\documentstyle[preprint,aps]{revtex} 
\draft
\begin{document}
\tighten
\title{Long-wavelength iteration scheme and scalar-tensor gravity}
\author{G.L. Comer}
\address{Department of Science and Mathematics,\\
Parks College of Saint Louis University,\\
Cahokia, IL 62206, USA}
\author{Nathalie Deruelle}
\address{D\'epartement d'Astrophysique Relativiste et de Cosmologie, \\
Centre National de la Recherche Scientifique,\\
Observatoire de Paris, 92195 Meudon Cedex, France\\
and Department of Applied Mathematics and Theoretical Physics,\\
University of Cambridge, Silver Street, \\
Cambridge CB3 9EW, England}
\author{David Langlois}
\address{D\'epartement d'Astrophysique Relativiste et de Cosmologie,\\
Centre National de la Recherche Scientifique,\\
Observatoire de Paris, 92195 Meudon Cedex, France}
\date{\today}
	
\maketitle
\begin{abstract}
Inhomogeneous and anisotropic cosmologies are modeled withing the framework
of scalar-tensor gravity theories. The inhomogeneities are calculated
to third-order in the so-called long-wavelength iteration scheme. We write
the solutions for general scalar coupling and discuss what happens to
the third-order terms when the scalar-tensor solution 
approaches  at first-order the general relativistic one. We work out in
some detail the case of Brans-Dicke coupling and determine the
conditions for which the anisotropy and inhomogeneity decay as time
increases. The matter is taken to be that of perfect fluid with a barotropic
equation of state.
\end{abstract}


\section{Introduction} \label{sec1}

Recently, Damour and Nordvedt \cite{ND} have shown quite generally that
General Relativity is an attractor of a large class of scalar-tensor
gravity theories in the context of cosmology (see also \cite{GQ}). Earlier,
Bekenstein and Meisels \cite{BM} had shown this to be true for Bekenstein's
variable mass theory \cite{B}. In both studies only Friedmann-Robertson-Walker
spacetimes were considered. We shall here extend these results to anisotropic
and inhomogeneous universes using the long-wavelength iteration scheme.
A general barotropic equation of state is used so that universes driven
by ordinary as well as ``inflationary'' matter can be considered.

The long-wavelength iteration scheme is an approach that allows for a
generic description of cosmological inhomogeneities and anisotropies
that goes beyond the linear regime. This approach was introduced by
Lifschitz and Khalatnikov \cite{LK} and Tomita \cite{T}. Recently various
groups have employed what amounts to the same scheme: Salopek, Stewart
and collaborators (see \cite{C} and references therein) use it in their
analysis of the Hamilton-Jacobi equation for general relativity; Comer,
Deruelle, Langlois and collaborators (see \cite{CDLP,DL} and references
therein, as well as \cite{DG} and \cite{TD}) discuss among other things
how the equation of state for matter affects the growth or decay of
cosmological inhomogeneities (see also \cite{TN,Ch,NT}); finally Soda 
{\it et al.} \cite{SII} solve the Hamilton-Jacobi equation in Brans-Dicke 
theory when matter is pressureless. We shall here extend the Soda 
{\it et al.} result to fluids with pressure and to other couplings, using 
the approach developped in \cite{CDLP} and \cite{DL}.

The essential point of the long-wavelength iteration scheme is that
spatial gradients must be small in the following sense: Choose a synchronous
gauge where the metric takes the form
\begin{equation}
 ds^2 = -dt^2 +\gamma_{ij} (t,x^k) dx^i dx^j \ . \label{1}
\end{equation}
Let $L$ be the characteristic comoving length on which the spatial metric
$\gamma_{ij}$ varies spatially: $\partial_i\gamma_{ik} \sim L^{-1}
\gamma_{jk}$. The Hubble time $H^{-1}$ (where $H\equiv \dot a/a$, $a^2\equiv
\gamma^{1/3}$, $\gamma$ being the determinant of $\gamma_{ij}$ and an
overdot signifying $\partial/\partial t$) is the characteristic time
on which the metric varies temporally: $\dot\gamma_{ij} \sim H\gamma_{ij}$.
The long-wavelength approximation, accurate when $L\gg R_H$ where $R_H\equiv
(aH)^{-1}$ is the local (comoving) Hubble radius, consists in neglecting 
all spatial gradients in the Einstein equations. However, the Hubble radius, 
in general, depends on time and eventually the condition $L\gg R_H$ can 
be violated.  To get more accurate (and ``longer-lived'') results, 
$\gamma_{ij}$ is expanded in terms of ever increasing spatial gradients of 
an arbitrary ``seed'' metric. Since each term in the expansion must be a 
tensor, the gradients of the ``seed'' enter via all possible contractions
and derivatives (leaving two free indices) of the Riemann (or equivalently
in 3-dimensions, Ricci) tensor formed from the ``seed''. The coefficients
of the gradient terms entering the expansion are time-dependent and are
determined order by order. For more discussion, see \cite{CDLP,DL}.

The only difficult part about implementing the iteration scheme is
finding exact solutions at the very first iteration. Fortunately Barrow
and Mimoso \cite{BMi} have given an algorithmic way of generating such
solutions for scalar-tensor gravity theories. We shall adopt their
techniques and extend them to the third-order. To get an explicit solution
however the scalar-tensor coupling must be specified. We shall work out
in some detail the Brans-Dicke coupling, consider the class
 studied by Damour and Nordvedt \cite{ND} and comment on
those chosen by Barrow and Mimoso \cite{BMi} and Serna and Alimi
\cite{SA}.  

In Sec. \ref{sec2} we give the action for scalar-tensor gravity in the
so-called ``Jordan frame" and write the field equations that follow in
a synchronous gauge. In Sec. \ref{sec3} we show how to solve them in
general in the long-wavelength approximation following Barrow and Mimoso
\cite{BMi} and particularize to the specific couplings mentioned above.
In Sec. \ref{sec4} the second iteration equations are written down and
their solutions explored in the late time limit. Finally the
corresponding calculations in the ``Einstein frame'' are outlined in an
Appendix.

The conclusion will be that, as in general relativity, inhomogeneities 
decay only if matter is inflationary.  However there is an additional 
condition on the scalar-tensor coupling which is it must be close 
enough to its general relativistic value (see eq. (\ref{42}) in the text).  

\section{The field equations} \label{sec2}

In the ``Jordan frame'' the action for scalar-tensor gravity is
\begin{equation}
  S = {1\over 16\pi} \int d^4x  \sqrt{-g} [ \phi R -\phi^{-1}
 \omega (\phi) g^{\mu\nu}  \partial_\mu \phi \partial_\nu
  \phi ] + S_M \ . \label{2}
\end{equation}
The function $\omega (\phi)$ is an a priori arbitrary function which
couples the scalar field $\phi$ to the metric $g_{\mu\nu}$. $S_M$ is the
action for the matter which will be taken to be a perfect fluid. Its
stress-energy tensor is
\begin{equation}
 T_{\mu\nu} = (\rho +p) u_\mu u_\nu + p g_{\mu\nu} \label{3}
\end{equation}
where $u_\mu$ is its 4-velocity $(u^\mu u_\mu =-1)$ and where the energy
density $\rho$ and pressure $p$ are chosen to be related by a barotropic
equation of state of the form
\begin{equation}
 p = (\Gamma -1) \rho \label{4}
\end{equation}
where $\Gamma$ is a constant ($0\leq \Gamma \leq 2$).

The field equations which follow from extremising eq. (\ref{2}) are
\begin{equation}
 R^\mu_\nu = {8\pi \over \phi} \left( T^\mu_\nu - {1\over 2} T\delta^\mu_\nu
  \right) + {1\over \phi} \left( \phi_{;\nu}^{;\mu}
    + {1\over 2} \phi_{;\sigma}^{;\sigma} \delta^\mu_\nu \right)
  + {\omega\over \phi^2} \partial_\nu \phi \partial^\mu \phi \label{5}
\end{equation}
and
\begin{equation}
 \Omega^{1/2} (\Omega^{1/2} \partial^\sigma \phi)_{;\sigma} = 8\pi T
  \label{6}
\end{equation}
where $\Omega \equiv 2\omega +3$ ($\Omega$ is assumed to be positive) 
and $T$ is the trace of the stress-energy tensor.

In the synchronous gauge (see eq. (\ref{1})), letting $\gamma^{ij}$ 
(which raises spatial indices) be the inverse of $\gamma_{ij}$ (which 
lowers spatial indices), defining $K_{ij} \equiv \dot\gamma_{ij}$,
$K\equiv K_i^i$, the field equations (\ref{5}-\ref{6}) become
\begin{equation}
  {1\over 2} \dot K + {1\over 4} K^j_i K^i_j = {8\pi\over \phi}
 \left( 1 -{3\over 2} \Gamma -\Gamma u^iu_i\right) \rho - {1\over 2\phi}
 \left( 3 \ddot \phi+{1\over 2} K \dot\phi -\phi^{|i}_{|i}\right)
    - {\omega\over \phi^2}\dot\phi^2 \ , \label{7}
\end{equation}
\begin{equation}
  -{1\over 2} (K^j_i -K\delta^j_i)_{|j}= {8\pi \over \phi} \Gamma
  \sqrt{1+u^ju_j}\, \rho u_i + {1\over 2\phi} (K^j_i \phi_{|j}
 -2\dot \phi_{|i})
  - {\omega\over \phi^2} \dot\phi \phi_{|i} \ , \label{8}
\end{equation}
\begin{eqnarray}
 {}^{(3)}R^j_i + {1\over 2\sqrt \gamma} {\partial \over \partial t} 
(\sqrt\gamma K^j_i) &=&
 {8\pi\over \phi} \left[ \Gamma u^ju_i + \left( {1-{\Gamma\over 2}}\right)
 \delta^j_i \right] \rho +{\omega\over \phi^2} \phi_{|i}\phi^{|j}+\nonumber\\
 &&+{1\over 2\phi} \left[ 2\phi^{|j}_{|i} - K^j_i \dot\phi - \left( \ddot\phi
  +{1\over 2} K\dot\phi -\phi_{|k}^{|k} \right) \delta^j_i \right] \ , 
\label{9}
\end{eqnarray}
\begin{equation}
- {\partial \over \partial t}(\Omega^{1/2}\dot\phi) -{1\over 2} 
K\Omega^{1/2}\dot\phi +(\Omega^{1/2}
  \phi^{|i})_{|i} = {8\pi\over \Omega^{1/2}} (3\Gamma -4)\rho  \label{10}
\end{equation}
where ${}^{(3)}R_{ij}$ is the 3-Ricci tensor formed out of $\gamma_{ij}$,
a slash means a covariant derivative with respect to $\gamma_{ij}$ and
$\gamma\equiv\det (\gamma_{ij})$. Eqs.~(\ref{7}-\ref{10}) are 11 equations
for the 11 unknowns $\rho$, $u_i$, $\phi$ and $\gamma_{ij}$.

\section{The long-wavelength field equations and their solutions}
\label{sec3}

In the long-wavelength approximation all spatial gradients in eqs.
(\ref{7}-\ref{10}) are ignored. Eq.~(\ref{8}) then simply implies that
\begin{equation}
  u_i = 0\ . \label{11}
\end{equation}
The trace-free part of eq. (\ref{9}) reduces to
\begin{equation}
 {\partial \over \partial t}\left[ \sqrt\gamma \phi \left( K^j_i - 
{1\over 3} K\delta^j_i \right)\right] = 0 \label{12}
\end{equation}
which yields
\begin{equation}
  K^j_i = {\dot A\over A} \delta^j_i + {1\over A^{3/2}\phi} S^j_i \label{13}
\end{equation}
where we have defined $A\equiv a^2 = \gamma^{1/3}$ so that $K=3\dot A/A$,
and where the so-called anisotropy matrix  $S^j_i$ is time-independent and  
traceless.  Eq.~(\ref{13}) can be integrated to yield the 3-metric 
$\gamma_{ij}$ (see \cite{CDLP,DL} for details). The final result is
\begin{equation}
 \gamma_{ij} = e^a_i \gamma_{ab} e^b_j \label{14}
\end{equation}
where the $e^a_i$, $a=1,2,3$, are the components of the 3 eigenvectors
of $S^j_i$ and $\gamma_{ab}= A\ {\rm Diag}\ [\lambda_1 e^{r_1\tau},
\lambda_2 e^{r_2\tau}, \lambda_3 e^{r_3\tau}]$ with $r_a$ the eigenvalues
of $S^j_i$ (with $r_1+r_2+r_3=0$ because $S^j_i$ is traceless), $\lambda_a$
three functions of space such that $\lambda_1\lambda_2\lambda_3 =1$, and
$d\tau /dt =(A^{3/2} \phi)^{-1}$. When $S^j_i=0$ the result simplifies
into
\begin{equation}
 \gamma_{ij}= A\, h_{ij} \label{15}
\end{equation}
where $h_{ij} (x^k)$ is an arbitrary ``seed'' metric.

The solution for $\rho$ is most easily obtained from the conservation
law $u^\nu T^\mu_{\nu;\mu} =0$ which follows from the field equations
(\ref{5}-\ref{6}) and boils down, at lowest order in the gradients, to
$\dot\rho +3\Gamma H\rho =0$ (with $H\equiv \dot A/2A$) and thus
\begin{equation}
  \rho = {3M_1\over 8\pi} A^{-3\Gamma/2} \label{16}
\end{equation}
where $M_1$ is a function of space only. Note that eq. (\ref{16}) holds 
whether $S^i_j$ is zero or not.

What remains is to solve for $A$ and $\phi$. They are given by the traces
of eq.~(\ref{9}), eq.~(\ref{7}) and eq.~(\ref{10}), neglecting all gradients
and using eqs. (\ref{13}) and (\ref{16}). (One of the equations of course 
is redundant and can be used as a consistency check.) When $S^i_j=0$ these
equations boil down to the Friedmann equations. Barrow and Mimoso
\cite{BMi} have shown how to reduce the problem to a quadrature when 
$\Omega(t)$ rather than $\Omega (\phi)$ is specified. For $S^i_j \not= 0$ 
Barrow and Mimoso's procedure can still be followed the only difference 
being that their constraint equation acquires an additional anisotropy term.
We therefore refer the reader to \cite{BM} for details. The final results
are
\begin{equation}
 \ln \phi = \int^\eta_0 d\eta {\eta\over g(\eta)}  \label{18}
\end{equation}
and
\begin{equation}
 A^{3(2-\Gamma)\over 2} = M_1 (4-3\Gamma) {g(\eta)\over \phi} \label{19}
\end{equation}
where $\eta$ is related to the time $t$ by
\begin{equation}
  d\eta /dt = A^{3(1-\Gamma)\over 2} \sqrt{3/\Omega}\ , \label{21}
\end{equation}
and where $g(\eta)$ is the function which characterizes the coupling through
the relation
\begin{equation}
 \Omega = {4-3\Gamma\over 3(2-\Gamma)^2}
  {[2g'-(3\Gamma-4)\eta]^2 \over 4g- (3\Gamma -4)\eta^2
  + S^j_i S^i_j / [6M^2_1 (4-3\Gamma)]} \label{20}
\end{equation}
with $g'\equiv dg/d\eta$.  Integration of this
equation for $g$ yields a constant of integration, which can be used
to parametrize the space of solutions. The anisotropy matrix only shows up 
in eq. (\ref{20}). However, as one can see from this equation, the anisotropy 
will be negligible at late times.

\subsection{Brans-Dicke coupling} \label{sec3a}

This case corresponds to $\Omega ={\rm const}$. The Friedmann solutions
($S^i_j=0$ in eq. (\ref{20})) have been extensively studied
(see, {\it e.g.}, \cite{BM,GFR,LP,HT,W} and references therein). 
From eq.~(\ref{20}) and the constancy of $\Omega$ we obtain
the ``generating'' function $g$:
\begin{equation}
 4 (4-3\Gamma) g(\eta) = [3\Omega (2-\Gamma)^2 -(4-3\Gamma)^2] \eta^2
 + \lambda \eta + {\lambda^2\over 12\Omega (2-\Gamma)^2}
 - {S^i_jS^j_i \over 6M^2_1} \label{22}
\end{equation}
where $\lambda$ is a function of space which parameterizes the solutions.
It is then an elementary exercise to
integrate eqs. (\ref{18}) and (\ref{19}). Only the relationship between
$\eta$ and $t$ [eq.~(\ref{21})] remains non trivial as it involves
a non elementary function of the type $\int d\eta (\eta-\eta_+)^\alpha
(\eta -\eta_-)^\beta$ where $\eta_{\pm}$ are the roots of $g(\eta)=0$.
At late time (or at all times if $\lambda =S^i_j =0$) the solution takes
the familiar form:
\begin{eqnarray}
 \phi \propto t^q \quad , \quad  &&q = {4(4-3\Gamma)\over (2-3\Gamma)
  (4-3\Gamma) +3\Gamma (2-\Gamma)\Omega} \ , \label{23} \\
 A \propto t^{2h} \quad , \quad  &&h = {2[\Omega(2-\Gamma)-(4-3\Gamma)]
   \over 
  (2-3\Gamma) (4-3\Gamma)+3\Gamma (2-\Gamma)\Omega} \ . \label{24}
\end{eqnarray}
(When $\Gamma=1$, eqs. (\ref{23}-\ref{24}) agree with the solution discussed
by Weinberg \cite{W}.)  As $A$ must remain positive we must have that 
$(4-3\Gamma) g(\eta) \geq 0$ (see eq.~(\ref{19})) and therefore
\begin{equation}
  \Omega \geq (4-3\Gamma)^2 / 3(2-\Gamma)^2 \ . \label{24bis}
\end{equation}
(See \cite{MW} for more discussion on the vanishing of anisotropy in
Brans-Dicke theory.)

\subsection{Dynamical couplings} \label{sec3b}

Various coupling functions $\Omega (\phi)$ have been considered in the
literature (see e.g. \cite{SA} and references therein). To be viable
the space of the corresponding cosmological solutions must exhibit a
basin of attraction towards general relativity. This means that when, say,
$\phi \to 1$ then $\Omega \to \infty$ (with $(d\Omega /d\phi) /\Omega^3\to 0$
\cite{N}). Many different behaviours of course are possible, for example
\begin{equation}
   \Omega \simeq \alpha / \ln \phi \label{25}
\end{equation}
which is the one considered by Damour and Nordvedt \cite{ND} and also by 
Barrow and Mimoso \cite{BMi} as it is the late-time behaviour of all 
solutions generated by a function $g\propto \eta^n$ with $n> 2$. As follows 
from eqs. (\ref{18}-\ref{21}) we have indeed \cite{BMi}
\begin{eqnarray}
  H &\equiv& \dot A/2A \simeq{ 2\over 3\Gamma t} \ , \label{26} \\
  P &\equiv& \dot\phi / \phi \simeq\ {\rm const}\ t^{-{2(n-2)(2-\Gamma)
    \over n\Gamma}-1} \ ,  \label{27} \\
 \Omega &\simeq& {4-3\Gamma\over 3(2-\Gamma)^2} {n^2\over (2-n)}
     {1\over \ln \phi} \ .  \label{28}
\end{eqnarray}
The alternative behaviour
\begin{equation}
  \Omega \simeq \lambda / (\phi - 1)^\varepsilon  \label{29}
\end{equation}
with ${1\over 2} <\varepsilon <2$ was studied by Serna and Alimi
\cite{SA} who analyzed numerically the solutions.

An important point to note is that it is not very meaningful
to look for solutions (either analytically \cite{BMi} or numerically
\cite{SA}) for $\Omega$ behaving as eq. (\ref{25}) or eq. (\ref{29}) at 
{\it all} times as the ``true'' function $\Omega$ may be significantly 
different from eq. (\ref{25}) or eq. (\ref{29}) away from the basin 
of attraction.

\section{The third-order solutions and their approach to general
relativity} \label{sec4}

The first-order being under control we now construct the third-order
solutions of the long-wavelength scheme describing large-scale
inhomogeneities. We shall set $S^i_j=0$ which in any case is negligible
at late time.

In accordance with the method presented in \cite{CDLP} the third-order
expansions for the 3-metric, the Brans-Dicke field, the energy density and
 the three-velocity are taken to be
\begin{equation}
 \gamma_{ij} =A(t) \left[ h_{ij} + f_2 (t) R_{ij} + {1\over 3}
  (g_2 (t) - f_2 (t)) R h_{ij} \right] \ , \label{30}
\end{equation}
\begin{equation}
 \phi = \phi (t) [ 1 + \phi_2 (t) R ] \ , \quad \rho=\rho(t)+\rho_2(t) R 
\ , \quad u_i=u_3(t)\nabla_i R \ , \label{31}
\end{equation}
where $R_{ij}$ is the Ricci tensor formed from the seed $h_{ij}$ and $R$
the corresponding Ricci scalar and where $A(t)$, $\phi(t)$ and $\rho(t)$ 
are the (Friedmann) first-order solutions of the previous section.

The procedure is to insert the above expansion in eqs.~(\ref{7})--(\ref{10})
and keep only terms with two spatial derivatives.  Gathering all ``like''
terms, {\it i.e.}, the coefficients of $R^j_i$ and $R$, gives the following
ordinary differential equations for $f_2$, $g_2$, and $\phi_2$:
\begin{eqnarray}
 \ddot f_2 + (3H+P) \dot f_2 &=& - {2\over A} \label{32} \ , \\
 \ddot g_2 + (2H+{P\over 2}) \dot g_2 &=& \Omega
     \left( {2-3\Gamma\over 4-3\Gamma} \right) X
   - [3\ddot \phi_2 +(2\Omega P+3H) \dot \phi_2 +P\dot\Omega \phi_2] 
\ , \label{33} \\
 \ddot g_2 + (6H+{5P\over 2}) \dot g_2+{2\over A} &=& 3\Omega
     \left( {2-\Gamma\over 4-3\Gamma} \right) X
   - 3[\ddot \phi_2 +(2P+5H) \dot \phi_2] \ , \label{34}
\end{eqnarray}
with
\begin{equation}
  X \equiv \ddot \phi_2 + \left( 2P +3H +{\dot\Omega\over \Omega}\right)
  \dot \phi_2 + {\dot\Omega\over 2\Omega P}
  \left( \dot P +2P^2 + PH +{P\ddot\Omega\over\dot\Omega}\right) \phi_2
  + {1\over 2} P\dot g_2 \ . \label{35}
\end{equation}
Eq.~(\ref{32}) can be integrated rightaway. Unfortunately the solutions
to the other two equations are not so easily forthcoming. However the
interesting regime is when $\Omega\to \infty$ (with $(d\Omega /d\phi)/
\Omega^3 \to 0$) for $t$ big which is the limit where the first-order
solutions approach their general relativistic limit. Now from eqs.
(\ref{18}-\ref{19}) it follows that $3(2-\Gamma) H/P = 2 (g'-\eta)/\eta$
and hence diverges as $\eta\to\infty$ when $\Omega\to\infty$, that is
when $g$ increases faster than $\eta^2$. In this limit
eqs.~(\ref{33}--\ref{34}) decouple, yielding
\begin{equation}
 \ddot g_2 + 3\Gamma H \dot g_2 \simeq {2-3\Gamma \over 2A}\ . \label{36}
\end{equation}
As for eq. (\ref{32}) it reduces to
\begin{equation}
 \ddot f_2 + 3H \dot f_2 \simeq -2/A \ . \label{37}
\end{equation}
Eqs.~(\ref{36}--\ref{37}) are the same as those obtained by Comer 
{\it et al.} (eq.~2.15 in \cite{CDLP}) in their study of inhomogeneities 
in general relativity. Thus, $f_2$ and $g_2$ ultimately have the same 
behaviour as the general relativity solutions discussed in \cite{CDLP}. 
Unfortunately, it is difficult to make precise statements about $\phi_2$ 
(and hence $\rho_2$ and $u_3$) even in the limit of diverging $\Omega$. We 
shall therefore turn to particular coupling functions.

\subsection{Brans-Dicke coupling}  \label{sec:4a}

The late time behaviour of all Friedmann solutions in the Brans-Dicke
theory of gravity is given by eqs. (\ref{23}-\ref{24}). 
Eqs.~(\ref{32}-\ref{35}) for the evolution of inhomogeneities can then 
be easily integrated yielding
\begin{equation}
 f_2 = \alpha_1 + \alpha_2 t^{1-3h-q} - {t^{2(1-h)}\over (1-h)(h+q+1)}
 \ , \label{38}
\end{equation}
\begin{equation}
 g_2 = \beta_1 + {\beta_2\over t} +{\Omega (2-3\Gamma) -3(4-3\Gamma)\over
  4\Omega (1-h)(3-2h)} t^{2(1-h)} \ , \label{39}
\end{equation}
\begin{eqnarray}
 \phi_2 &=& \gamma_1 +\gamma_2 t^{1-3h-q} + {4-3\Gamma\over
    3[\Omega (2-\Gamma) -(4-3\Gamma)]} {\beta_2 \over t}\nonumber\\
& & +{(4-3\Gamma)t^{2(1-h)}\over 4\Omega (3-2h)(1-h)}
 \left[ {(4-3\Gamma)^2-\Omega (3\Gamma^2-12\Gamma+4)\over
  (4-3\Gamma)^2 +\Omega (2-\Gamma) (2+3\Gamma)} \right] \ , \label{40}
\end{eqnarray}
where $\alpha_1$, $\alpha_2$, $\beta_1$, $\beta_2$, $\gamma_1$, $\gamma_2$
are integration constants. One can check that in the limit $\Omega\to\infty$
the results for $f_2$ and $g_2$ agree with those of Comer {\it et al.}
\cite{CDLP}. Furthermore, when $\Gamma=1$, corresponding to a dust fluid,
 all three agree with the solution of Soda {\it et al.} \cite{SII}.

The growth or decay of $f_2$, $g_2$ and $\phi_2$ depends on the signs of the
exponents in eqs. (\ref{38}-\ref{40}). Recalling the condition of positivity
of $A$ (eq.~(\ref{24bis})) one finds that $(f_2, g_2,\phi_2)$ all decay as 
time increases if and only if
\begin{equation}
 \Gamma < 2/3 \ , \label{41}
\end{equation}
that is if matter is of an inflationary type (the same condition for
decay holds in general relativity, see \cite{CDLP}), and if
\begin{equation}
 \Omega > {(4-3\Gamma)^2\over (2-\Gamma)(2-3\Gamma)} \ . \label{42}
\end{equation}
This condition on $\Omega$ is not demanding if $p=-\rho$ $(\Gamma=0)$
but becomes all the more stringent as $\Gamma \to 2/3$.

As for the three-velocity its time-dependence is obtained from
\begin{equation}
{8\pi\Gamma\rho\over \phi}u_3={1\over 3}\dot g_2-{1\over 12}\dot f_2
-\left[H+P\left({1-\Gamma\over 2}\right)\right]+\dot \phi_2 \ ,
\end{equation}
which yields $u_3 \propto t^{1+q+h(3\Gamma -2)}\rightarrow
t^{3-4/3\Gamma}$ when $\Omega\rightarrow\infty$ in accordance 
with the general relativity results \cite{CDLP}, unless $\Gamma=1$ (dust)
in which case $u_3=0$ (since dust can be comoving in a synchronous 
gauge).  Finally the time behaviour of the energy density correction 
$\rho_2$ is given by $\rho_2\propto t^{q+h(3\Gamma-2)}\rightarrow 
t^{2(1-2/3\Gamma)}$, again in agreement with the general relativity 
result.

\subsection{Dynamical coupling: an example}  \label{sec:4b}

Let us consider as an example the dynamical coupling generated by the
function $g=E\eta^n$ \cite{BMi,ND}. The late time behaviour of the 
long-wavelength, Friedmann solution, is given in eqs. (\ref{26}-\ref{28}).
As already mentionned, as $\Omega\to\infty$ in this limit, $P/H\to 0$ so
that the functions $f_2$ and $g_2$  are governed by the same equation as
in general relativity, eqs.~(\ref{36}-\ref{37}), the general solutions of
which are
\begin{equation}
 f_2 = \alpha_1 +\alpha_2 t^{-(2-\Gamma) /\Gamma}
       - {9\Gamma^2\over 9\Gamma^2-4} t^{2-4/3\Gamma} \label{43}
\end{equation}
and
\begin{equation}
 g_2 = \beta_1 +\beta_2 t^{-1}  
       -{9\Gamma^2\over 4(9\Gamma-4)} t^{2-4/3\Gamma} \ . \label{44}
\end{equation}
As for the equation for $\phi_2$ [eqs.~(\ref{33}-\ref{35})] it can be 
integrated and yields
\begin{eqnarray}
 \phi_2 &=&\gamma_1 t^{-4(n-2)(2-\Gamma)\over 3\Gamma n}
  + \gamma_2 t^{-2(2-\Gamma)\over 3\Gamma}
  - {(4-3\Gamma)[2(2-\Gamma)-n]\beta_2 t^{-{2(2-\Gamma)(n-2)\over \Gamma n}-1}
             \over 6(2-\Gamma)[(2\Gamma-3)n+2(2-\Gamma)]}
  +\nonumber \\
 && {9\Gamma^2n (4-3\Gamma)[3\Gamma(2-\Gamma)+n(3\Gamma-2)]\over
   4(9\Gamma-4) (2-\Gamma)(3\Gamma-2) [(9\Gamma -10)n +12 (2-\Gamma)]}
   t^{-4[(4-3\Gamma)n-3(2-\Gamma)]\over 3n\Gamma}\ . \label{45}
\end{eqnarray}
All terms decay if $\Gamma < 2/3$, that is for inflationary matter. However
$\phi_2$ can still decay if ${2\over 3} < \Gamma < {2(2n-3)\over 3(n-1)}< 4/3$.

\def\a{\alpha}
\def\beq{\begin{equation}}
\def\eeq{\end{equation}}
\def\d{\delta}
\def\hf{{\scriptstyle{1\over 2}}}
\def\td{{\scriptstyle{1\over3}}}
\def\fh{{\scriptstyle{1\over4}}}
\def\g{\tilde g}
\def\G{\Gamma}

\section{ Appendix: Long-wavelength scheme in the Einstein frame}

In the Jordan frame, the action is given by 
\begin{equation}
S={1\over 16\pi}\int d^4x\sqrt{-\g}\left[\phi{\tilde R}
-\phi^{-1}\omega(\phi)\g^{\mu\nu}\partial_\mu\phi\partial_\nu\phi\right]
+S_m \ .
\eeq
In this appendix, contrarily to the main text, all the physical quantities 
will be labelled by a tilde, whereas the non-physical quantities will remain 
bare.  Upon using a change of the metric and of the fields defined by
\begin{equation}
\g_{\mu\nu}=e^{2a(\varphi)}g_{\mu\nu} \ ,
\eeq
\begin{equation}
\phi^{-1}= e^{2a(\varphi)}
\eeq
and 
\begin{equation}
\alpha^2=(2\omega +3)^{-1} \ ,
\eeq
with $\alpha(\varphi)={\partial a/\partial\varphi}$,
one can reexpress the above action in the form
\begin{equation}
S={1\over 16\pi }\int d^4x\sqrt{-g}\left[ R
-2 g^{\mu\nu}\partial_\mu\varphi\partial_\nu\varphi\right]
+S_M[\psi_m, e^{2a(\varphi)} g_{\mu\nu}] \ .
\eeq
This form is said to correspond to the ``Einstein frame'' because
the curvature term in this new action is the same as in the Einstein-Hilbert
action for general relativity.  The advantage of this formulation is that 
most of the equations can be obtained almost directly from the equations 
of \cite{CDLP} by combining the results for a perfect fluid and for a 
scalar field.

The Brans-Dicke theory corresponds to a ``linear'' coupling in the sense that
\begin{equation}
a(\varphi)=\alpha\varphi \ .
\eeq
Then 
\begin{equation}
\phi=e^{-2\alpha\varphi} \ .
\eeq

The field equations in the Einstein frame read
\begin{equation}
R_{\mu\nu}=2\partial_\mu\varphi \partial_\nu\varphi+\kappa(T_{\mu\nu}-\hf
Tg_{\mu\nu})\ , \eeq
and
\begin{equation}
\nabla_\mu\nabla^\mu\varphi=-{\kappa\over 2}\alpha(\varphi)T \ ,
\eeq
with $\kappa\equiv 8\pi$.  In the case of a perfect fluid, this gives
\begin{equation}
R_{\mu\nu}=2\partial_\mu\varphi \partial_\nu\varphi+\kappa(\rho+p)u_\mu u_\nu
+{\kappa\over 2}(\rho-p)g_{\mu\nu} \ ,
\eeq
\begin{equation}
\nabla_\mu\nabla^\mu\varphi=-{\kappa\over 2}\alpha(\varphi)(3p-\rho) \ .
\eeq
One must however be aware that the $\rho$ and $p$ that appear here, 
defined by $T^{\mu\nu}=(\rho+p)u^\mu u^\nu + pg^{\mu\nu}$, with 
$g_{\mu\nu}u^\mu u^\nu=-1$, are not the physical energy density and pressure,
but are related to the latter by $\rho=e^{4a}\tilde \rho$, $p=e^{4a}\tilde p$.

In a synchronous gauge (adopting all the notations of the main text
but for the {\it unphysical} metric)
\begin{equation}
 {}^{(3)}R^j_i+{1\over 2\sqrt{\gamma}}{\partial\over\partial t}
  \left(\sqrt{\gamma}K_i^j\right)={\kappa\over 2}\rho
  \left[2\Gamma u^ju_i+\delta^i_j(2-\Gamma)\right]
+2\partial_i\varphi\partial^j\varphi \ ,
\label{a}
\eeq
\begin{equation}
{1\over 2}\dot K+{1\over 4 }K^i_jK^j_i=\kappa\rho\left[1-{3\over 2}\G
-\G u_ku^k\right]-2\dot\varphi^2 \ ,
\label{b}
\eeq
\begin{equation}
-{1\over 2}\left(K^j_{i;j}-K_{,i}\right)= \kappa\rho\Gamma u_i\sqrt{1+u^ku_k}
-2\dot\varphi\partial_i\varphi \ ,
\label{c}
\eeq
and
\begin{equation}
\ddot\varphi+{1\over 2}K\dot\varphi-\nabla_i\nabla^i\varphi={\kappa\over 2
}\alpha (3p-\rho) \ .
\label{d}
\eeq

At first-order the trace of eq. (\ref{a}) yields 
\begin{equation}
A^{-3/2}{\partial\over\partial t}\left(A^{3/2}{\dot A\over A}\right)=
\kappa\rho(2-\G) \ . \label{a0}
\eeq
Using this equation, eq. (\ref{b}) at first-order then gives 
the following equation for $A$:
\begin{equation}
     {\partial\over\partial t}\left({\dot A\over A}\right)+{3\Gamma\over 4}
\left({\dot A\over A}\right)^2+{2-\Gamma\over 8} S^i_jS^j_i A^{-3}
+(2-\G)\dot\varphi^2=0 \ .
\end{equation}
In contrast with pure general relativity, we now have an additional term 
coming from the scalar field. One thus needs an extra equation to close 
the system and solve for $A$. This is of course given by eq. 
(\ref{d}), which reads at first-order 
\begin{equation}
\ddot\varphi+3H\dot\varphi={\kappa\over 2}\alpha(3\G-4)\rho \ .
\eeq
Combining with eq. (\ref{a0}), one finds
\begin{equation}
{\partial\over\partial t}\left(A^{3/2}{\dot A\over A}\right)=
{2(2-\G)\over\alpha (3\G-4)}{\partial\over\partial
t}\left(A^{3/2}\dot\varphi\right) \ .
\eeq

Let us now consider the Brans-Dicke case, {\it i.e.}, when $\alpha$ is 
a constant.  Then the previous equation can be integrated immediately, so 
as to give
\begin{equation}
{\dot A\over A}-{2(2-\G)\over\alpha (3\G-4)}\dot\varphi=
{{\cal C}\over A^{3/2}} \ .
\eeq 
In the simplest case ${\cal C}=0$ and $S^i_jS^j_i=0$, one finds
\begin{equation}
{\dot A\over A}=2\lambda t^{-1} \ , \qquad \lambda={2(2-\G)\over
 3\G(2-\G)+\alpha^2(3\G-4)^2} \ ,
\eeq
\begin{equation}
\dot\varphi=\xi t^{-1} \ , \qquad \xi={2\alpha(3\G-4)\over 
 3\G(2-\G)+\alpha^2(3\G-4)^2} \ .
\eeq

Let us make explicit the relation between this result and eqs.  
(\ref{23}-\ref{24}) in the main body of this article.  The relation 
between the ``Jordan time'' $\tilde t$ and the ``Einstein time''
$t$ is given by 
\begin{equation}
\tilde t\sim t^{\alpha\xi +1} \ . \label{time}
\eeq
Hence 
\begin{equation}
\tilde A\sim {\tilde t}^{2\alpha\xi+2\lambda\over \alpha\xi+1} \ ,
\eeq
which agrees with eq. (\ref{23}), using $\alpha^2\equiv \Omega^{-1}$. 
Finally,
\begin{equation}
\phi\sim \tilde t^{-{2\alpha\xi\over \alpha\xi+1}} \ ,
\eeq
which agrees with eq. (\ref{22}).

The third-order is obtained by taking into account the terms
neglected previously, in particular the Ricci tensor of the three-metric.
It is consistent with the approximation to take the Ricci tensor of
${}^{(1)}\gamma_{ij}$ in place of ${}^{(3)}R_i^j$, and convenient to
write it in terms of the Ricci tensor (denoted $R_{ij}$) of $h_{ij}$,
seen as the components of a metric, to be called ``seed" metric.  We
will thus look for corrections to the metric of the form
\begin{equation}
 {}^{(3)}\gamma_{ij}=
A(t)\left\{f_2(t) R_{ij}+{1\over 3}\left[g_2(t)-f_2(t)\right] R h_{ij}
\right\} \ ,   \label{eq:15}
\end{equation}
where $R_i^j\equiv h^{ik}R_{kj}$ and $R\equiv R_i^i$ and corrections
to the matter of the form
\begin{equation}
\varphi=\varphi(t)+\varphi_2(t) R \ ,\qquad \rho=\rho(t)+\rho_2(t)R \ .
\end{equation}
We substitute $\gamma_{ij}= {}^{(1)}\gamma_{ij}+{}^{(3)}\gamma_{ij}$ in 
the Einstein equation and keep the terms with two gradients.
The traceless part of eq. (\ref{a}) gives an equation for $f_2$,
\begin{equation}
\ddot f_2+{3\over 2}{\dot A\over A}\dot f_2=-{2\over A} \ .
\label{f2}
\eeq
Eq. (\ref{b}) then yields 
\begin{equation}
{\kappa\rho_2}={1\over 2-3\G}\left(\ddot g_2+2H\dot g_2+8\dot\varphi_o
\dot\varphi_2\right) \ . \label{rho2}
\eeq
Using this relation, the trace part of eq. (\ref{a}) gives an equation 
involving $g_2$ and $\varphi_2$:
\begin{equation}
\ddot g_2+{3\over 2}\Gamma{\dot A\over A}\dot g_2
+6(2-\G)\dot\varphi\dot\varphi_2={2-3\Gamma\over 2A}.
\label{g2}
\eeq
Eq. (\ref{d}) provides, once more after use of eq. (\ref{rho2}), another
such equation:
\begin{equation}
\ddot\varphi_2+3H\dot\varphi_2+{1\over 2}\dot\varphi{\dot g_2}=
{3\G-4\over 2-3\G}\alpha\left[{1\over 2}\ddot g_2 +H\dot g_2 +4\dot\varphi
\dot\varphi_2\right] \ .
\eeq

In the case of Brans-Dicke the system of equations can be solved explicitly.
All the particular solutions for $f_2$, $g_2$ and $\varphi_2$ behave
like $t^{2-2\lambda}$. After transformation of the time according to 
eq. (\ref{time}), one sees that this is in agreement with the results 
in the ``Jordan'' frame.

\section{Acknowledgements}

We wish to thank J. D. Bekenstein for providing references on scalar-tensor 
gravity, and the variable mass theory in particular.
One of us (GLC) gratefully acknowledges a 1995 NASA/ASEE Summer Faculty
Fellowship with the Jet Propulsion Laboratory of the California Institute of
Technology and Parks College of Saint Louis University for partial support 
as well as the hospitality of Observatoire de Paris where part of this 
research was carried out.

\end{document}